\documentclass{elsart}
\usepackage{graphicx}
\usepackage{amssymb}
\newcommand{\beq}{\begin{eqnarray}}
\newcommand{\eeq}{\end{eqnarray}}

\begin{document}

\begin{frontmatter}

\title{The string of variable density: perturbative and non-perturbative results}
\author{Paolo Amore}
\ead{paolo.amore@gmail.com}
\address{Facultad de Ciencias, CUICBAS, Universidad de Colima,\\
Bernal D\'{i}az del Castillo 340, Colima, Colima, Mexico}

\begin{abstract}

We obtain systematic approximations for the modes of vibration of a string of variable density, which
is held fixed at its ends. These approximations are obtained iteratively applying three theorems which are
proved in the paper and which hold regardless of the inhomogeneity of the string.
Working on specific examples we obtain very accurate approximations which are compared both with the 
results of WKB method and with the numerical results obtained with a collocation approach. Finally, we
show that the asymptotic behaviour of the energies of the string obtained with perturbation theory, 
worked to second order in the inhomogeinities, agrees with that obtained with the WKB method and 
implies a different functional dependence on the density that in two and higher dimensions.
\end{abstract}

\begin{keyword}
{Helmholtz equation; inhomogeneous string; perturbation theory; collocation method}
\end{keyword}

\end{frontmatter}

\section{Introduction}
\label{intro}

We consider the problem of describing the vibrations of a string of variable density, which is held fixed 
at its ends (Dirichlet boundary conditions). This problem has been investigated in depth in a series of papers by 
Krein, who has considered both the direct and inverse problem~\cite{Krein51,Krein52a,Krein52b,Krein53,Krein54}. 
In recent years, Beals, Sattinger and Szmigielsky \cite{BSS07},
have studied the string density problem in connection with the Camassa-Holmes equation for shallow water waves. 

Although the problem for a string is considerably simpler than the corresponding two dimensional version, i.e. an inhomogeneous membrane 
of arbitrary density, exact solutions are known only in few cases: two examples of exactly solvable problems are the 
strings studied by Lord Rayleigh \cite{Rayleigh45} and in the late $40$s by Borg \cite{Borg46}, which correspond
to case where the reciprocal of the densities vary as the fourth or the second power of the position respectively. 
Other examples of exactly solvable inhomogeneous strings are given by Horgan and Chang \cite{Horgan99}.

Recently Gottlieb \cite{Gottlieb02} has studied these examples and he has obtained a transformation which maps the original
variable density string into a family of strings which are isospectral to it: in this way Gottlieb shows that 
the case discussed by Borg can be obtained from a homogeneous string via his transformation.
Another case of problem which can be formally solved exactly is a string  of density which depends linearly on the position:
Fulcher \cite{Fulcher85} has obtained the exact solutions for this case, in terms of a complicated trascendental equation. 

In the cases which cannot be solved exactly or where the exact solution is of no practical use, as for the example discussed by Fulcher, 
one needs to resort to approximations: a typical approach is the WKB method, which correctly describes the leading behaviour of 
the highest part of the spectrum, although it is less precise for the lowest excited modes 
\footnote{The application of the WKB method to the variable density string is discussed in detail by Bender and Orszag in their book 
\cite{BO78}.}. Using this approach, Crawford has obtained a simple analytical formula for the problem
solved by Fulcher,  which describes all the spectrum with an accuracy of few percent \cite{Crawford87}.
An example of application of the WKB method beyond the leading order to a nontrivial problem of a density with a rapidly 
oscillatory behaviour is discussed by Castro and Zuazua \cite{CZ00a,CZ00b}.
An interesting computational scheme for the solution of the inhomgeneous one dimensional Helmholtz equation has been recently discussed
by Rawitscher and Liss in ref.~\cite{RL10}, using a spectral expansion in term of Chebyshev polynomials.

An alternative to the WKB method is provided by perturbation theory (PT), which however is bounded to 
cases where the density is only slightly perturbed with respect to the density of an exactly solvable problem.
Using this approach in the case of two dimensional membranes (or billiards), and performing a suitable
resummation of the perturbative terms, it has been possible to derive Weyl's law, which relates the spectrum
of the billiard to its area (see \cite{Amore10b}). In the present paper we show an interesting (and unexpected) result: 
the Weyl's law for a one dimensional string corresponds to a {\sl different functional} of the density 
than in higher dimensions (the problem of a 
homogeneous membrane  in two dimensions is isospectral to the problem of an inhomogeneous membrane with density obtained 
from the conformal  map which send the border of the membrane to a reference border). The results that we obtain here, 
which are exact to second order, agree with the general WKB formula and provide a link between these two different
approaches. 

The central results of the present paper  are contained in three theorems, which provide an explicit iterative 
tool to build increasingly accurate approximations to the modes of the strings, without being restricted
to slightly inhomogeneous systems or highly excited states. We discuss the application of these theorems to specific 
examples. The comparison of these results with the precise numerical results  obtained with a collocation method allows
to verify explicitly their convergence.

The paper is organized as follows: in Section \ref{collocation} we describe the collocation approach; 
in Section \ref{nonperturbative} we enunciate and prove the theorems; in Section \ref{perturbation} we
discuss the application of Perturbation Theory and prove that, to second order, one recovers the results of WKB method;
in Section \ref{applications} we discuss two applications of the general results obtained in the previous sections;
finally, in Section \ref{conclusions} we draw our conclusions.

\section{Collocation approach}
\label{collocation}

In this section we briefly describe a collocation approach which can be used to solve the Helmholtz
equation for a one dimensional string with variable density. The method that we are using is the 
{\sl Conformal collocation method} (CCM) that we have devised in Ref.~\cite{Amore08}, where we have shown that
it can be used to obtain precise numerical approximations to the energies and wave functions of the
Helmholtz equation on an arbitrary two dimensional region with Dirichlet boundary conditions. 
This method has also been used recently to obtain very precise numerical estimates for arbitrary domains 
in the plane, either simply connected, \cite{Amore10,Amore10b}, or with a hole, \cite{AA10},
these last domains being known as "quantum rings". 

In order to make our discussion self contained we briefly review the main features of the method.
Our starting point is the inhomogeneous Helmholtz equation on the interval $[-L,L]$:
\beq
- \frac{d^2}{dx^2} \psi_n(x) = E_n \rho(x) \psi_n(x) \ ,
\label{helm1d}
\eeq
where $\rho(x) > 0$ is the density of the string and $\psi_n(\pm L) = 0$. The eigenfunctions 
$\psi_n(x)$ are orthogonal with respect to the weight function $\rho(x)$:
\beq
\int_{-L}^{L} \psi_n(x) \psi_m(x) \rho(x) dx = \delta_{nm}  \ .
\label{ortho1d}
\eeq

In order to discretize this eigenvalue equation we introduce a set of functions, which we 
call {\sl Little Sinc Functions} (LSF), discussed in \cite{Amore07} and fullfilling Dirichlet 
boundary conditions~\footnote{In \cite{Amore09} we have also discussed three other sets of LSF
obeying different boundary conditions: choosing one of these alternative sets would then correspond
to solve Helmholtz equation (\ref{helm1d}) with the corresponding boundary condition.}:
\beq
s_k(N,L,x) = \frac{(-1)^k}{N}
\frac{\cos\left(\frac{\pi k}{N}\right) \sin\left(\frac{N\pi x}{2L}\right)}
{\sin \left(\frac{\pi x}{2L}\right) - \sin \left(\frac{\pi k}{N}\right)}.
\label{LSF}
\eeq
with $k=-N/2+1, \dots, N/2-1$. These functions define an homogeneous grid 
$\displaystyle x_k = \frac{2Lk}{N}$ and obey the orthogonality relation
\beq
\int_{-L}^{+L} s_k(N,L,x) s_j(N,L,x) dx = h \delta_{kj} ,
\eeq
where $h \equiv \frac{2L}{N}$ is the spacing of this grid.

A function $f(x)$ obeying Dirichlet boundary conditions may be interpolated using 
the $s_k(h,N,x)$ as
\beq
f(x)  \approx \sum_{k=-N/2+1}^{N/2-1} f(x_k) s_k(h,N,x) \ .
\label{B2}
\eeq

To better understand the validity of this expression we recall the definition of LSF
\beq
s_k(N,L,x) = \frac{2L}{N} \sum_{n=1}^N \psi_n(x) \psi_n(x_k) \ ,
\label{LSF1}
\eeq
from which eq.~(\ref{LSF}) was obtained in ref.\cite{Amore07}. 
Here $\psi_n(x) = \frac{1}{\sqrt{L}} \sin \frac{n\pi (x+L)}{2L}$ are the 
solutions for a homogeneous string of unit density, which form a basis.

Substituting this expression inside the rhs of eq.~(\ref{B2}), we obtain
\beq
\sum_{n=1}^N \left[\frac{2L}{N}\sum_{k⁼-N/2+1}^{N/2-1} f(x_k) \psi_n(x_k)\right] \ \psi_n(x) \ .
\eeq

One may easily recognize that the term in parenthesis is a Riemann sum and that, for $N\rightarrow \infty$, 
it converts to the integral $\int_{-L}^{+L} f(y) \psi_n(y)dy$, thus leading to the standard decomposition
of a function $f(x)$ in the basis $\left\{ \psi_n(x) \right\}$. This discussion can also be found
in Ref.~\cite{Amore07}.

In the same way one may obtain an interpolation formula for the second derivative of the function, simply
deriving twice the expression above and thus obtain
\beq
\frac{d^2f(x)}{dx^2}  &\approx& \sum_{k=-N/2+1}^{N/2-1} f(x_k) \ \frac{d^2s_k(x)}{dx^2} \nonumber \\
&\approx& \sum_{k=-N/2+1}^{N/2-1} \sum_{j=-N/2+1}^{N/2-1} f(x_k) \ \left. \frac{d^2s_k(x)}{dx^2}\right|_{x_j} 
s_j(h,N,x) \nonumber \\
&\equiv&  \sum_{k=-N/2+1}^{N/2-1} \sum_{j=-N/2+1}^{N/2-1} f(x_k) \ c_{kj}^{(2)} \ 
s_j(h,N,x)  , 
\label{B3}
\eeq
where the matrix of elements $c_{kj}^{(2)} \equiv \left. \frac{d^2s_k(x)}{dx^2}\right|_{x_j}$
provides a representation for the second derivative operator on the grid.

After writing eq.~(\ref{helm1d}) into the equivalent form
\beq
- \frac{1}{\rho(x)} \frac{d^2}{dx^2} \psi(x) = E \psi(x) \ ,
\label{helm1db}
\eeq
one may obtain a representation of the differential operator on the left, $\hat{O} = -\frac{1}{\rho(x)} \frac{d^2}{dx^2}$,
in terms of the LSF introduced above as: for a set of LSF with a given $N$, the matrix elements of $\hat{O}$
on the grid are given by
\beq
O_{kk'} = -  \frac{1}{\rho(x_k)}  c_{kk'}^{(2)} 
\eeq
where $k,k'= -N/2+1, \dots , N/2-1$. 

As pointed out in \cite{Amore10} this form of $\hat{O}$ is not manifestly hermitean, so that it is more convenient
to work with the symmetrized form of the operator $\hat{O}_{sym} = - \frac{1}{\sqrt{\rho(x)}} \frac{d^2}{dx^2}
\frac{1}{\sqrt{\rho(x)}}$, whose matrix elements read
\beq
O^{sym}_{kk'} = -  \frac{1}{\sqrt{\rho(x_k)}}  c_{kk'}^{(2)} \frac{1}{\sqrt{\rho(x_{k'})}}  \ .
\eeq

The only differences in this discussion from the cases discussed in \cite{Amore10b,Amore08,Amore10,AA10} are that 
we are limiting ourselves to one dimension and that $\rho(x)$ is 
a {\sl physical density}, i.e. cannot be considered as obtained from a conformal mapping.

The reader may be worried that dealing with the symmetrized operator may not be equivalent to dealing with the original
operator: we may easily convince ourselves that this is not so. Let us consider the Helmholtz equation (\ref{helm1db})
and introduce the function $\phi(x) \equiv \sqrt{\rho(x)} \psi(x)$. If we substitute this function in the equation we obtain
\beq
- \frac{1}{\sqrt{\rho(x)}} \frac{d^2}{dx^2} \frac{1}{\sqrt{\rho(x)}} \phi(x) = E \phi(x) \ ,
\label{helm1db2}
\eeq
which involves the symmetrized operator introduced earlier. In essence the eigenvalues of $\hat{O}$ and $\hat{O}_{sym}$ are
the same, while the eigenfunctions are related by a factor $\sqrt{\rho(x)}$: this also implies that the eigenfunctions
of eq.(\ref{helm1db}) are orthogonal with respect to the weight $\rho(x)$, as stated in eq.~(\ref{ortho1d}).

The advantages of the present collocation approach are clear:
\begin{itemize}
\item the representation on the grid of the Helmholtz equation with variable density is obtained straightforwardly;
\item the matrix representing the differential operator is the product of a universal matrix (the matrix for the laplacian
on a finite interval) with the matrix (or matrices depending if one is using the symmetrized form of the operator or not)
which depends on the density but is {\sl diagonal};
\item the calculation of the universal matrix corresponding to a given grid may be done once and for all,
whereas the diagonal density matrices need to be calculated for each specific density; clearly the computational
price of this second operation is negligible with respect to the first task. The correct computational strategy
can therefore be calculating the matrix of the 1d laplacian for different grids and then store them for later use;
\item at no time has one to perform numerical integrations;
\item the collocation method provides {\sl exact} results for constant densities: for this reason its application
to problems with slightly inhomogeneous densities provides very precise results;
\end{itemize}

After this process has been carried out one obtains a $(N-1) \times (N-1)$ hermitean matrix whose
eigenvalues and eigenvectors will provide approximations to the lowest $N-1$ energies and wave functions
of eq.~(\ref{helm1d}).

\section{Non-perturbative approach}
\label{nonperturbative}

In this section we wish to obtain explicit formulas for the states of a string of variable density which 
obey eq.~(\ref{helm1d}). These formulas are non-perturbative, i.e. they do not depend 
on the inhomogeneities being small.

We state the following theorem:

\begin{thm}
\label{theo1}
Let $\xi_0(x)$ be an arbitrary function defined on the interval $(-L,L)$ (being $\xi_0(x)$ arbitrary we assume that it
has nonzero overlap with the true fundamental solution). For $n\rightarrow \infty$ the sequence of functions 
\beq
\xi_n(x) \equiv \sqrt{\rho(x)} \int_{-L}^x dy \left[ \kappa_n - \int_{-L}^y dz \sqrt{\rho(z)} \xi_{n-1}(z) \right] \ ,
\label{iter}
\eeq
with $\kappa_n =  \frac{1}{2L} \int_{-L}^{L} dy \int_{-L}^{y} dz \sqrt{\rho(z)} \xi_{n-1}(z)$, converges to the fundamental 
mode of the Helmholtz equation (\ref{helm1d}) with Dirichlet boundary conditions, $\xi_n(\pm L) = 0$:
\beq
\Psi_0(x) = \lim_{n\rightarrow \infty} \tilde{\xi}_n(x) \ ,
\eeq
where $\tilde{\xi}_n(x) \equiv \frac{\xi_n(x)}{\sqrt{\rho(x)}}$.
\end{thm}

\begin{pf*}{Proof}
Since $\xi_0(x)$ is an arbitrary function of $x$, it will have a nonzero overlap with the exact wave function
of the ground state, $\Psi_0(x)$. The operator 
$\hat{O} = - \frac{1}{\sqrt{\rho}} \ \frac{d^2}{dx^2} \  \frac{1}{\sqrt{\rho}}$ associated to the 
eq.~(\ref{helm1d}) has a spectrum which is bounded from below, positive and simple (Theorem 2.1 of \cite{BSS07}).
The spectrum is not bounded from above. The inverse operator $\hat{O}^{-1}$ therefore has a spectrum
which is also positive and simple, but it is now bounded both from below and from above. The function
obtained by applying $\hat{O}^{-1}$ to $\xi_0(x)$, corresponding to the case $n=1$ in eq.~(\ref{iter}) obeys
Dirichlet boundary conditions and it has a larger overlap with the exact wave function 
$\Psi_0(x)$~\footnote{Recall the orthogonality relation (\ref{ortho1d}).}:
\beq
\frac{\left| \int_{-L}^{+L} \sqrt{\rho(x)} \Psi_0(x) \xi_1(x) dx \right|}{\sqrt{\int_{-L}^{+L} \xi_1(x)^2 dx}} 
\geq \frac{\left| \int_{-L}^{+L} \sqrt{\rho(x)} \Psi_0(x) \xi_0(x) dx \right|}{\sqrt{\int_{-L}^{+L} \xi_0(x)^2 dx}}  \ .
\eeq
As a matter of fact the operator $\hat{O}^{-1}$ acting on $\xi_0(x)$
"inflates" the components of $\xi_0(x)$ corresponding to the lowest energy states of $\hat{O}$. The iteration
of this precedure will then provide a sequence of functions which converge to $\Psi_0(x) \sqrt{\rho(x)}$ as 
$n\rightarrow \infty$. In the case that $\xi_0(x)$ is orthogonal to the first $k$ states of eq.~(\ref{helm1d}), 
$\Psi_0(x)\sqrt{\rho(x)}, \dots, \Psi_{k-1}(x)\sqrt{\rho(x)}$, then the sequence of functions $\xi_n(x)$ 
will converge to $\Psi_k(x)\sqrt{\rho(x)}$.
\end{pf*}

Incidentally, one may consider this theorem (and the more general Theorem 1 of \cite{Amore10}) as the 
generalization of the celebrated {\sl power method} of linear algebra to operators in a Hilbert space.
Notice also that eq.~(\ref{iter}), where $\xi_n(x) = \xi_{n-1}(x) = \xi(x)$, corresponds to Helmholtz equation in 
its integral form.

We may easily extend this theorem to calculate the first $N$ states of the inhomogeneous Helmholtz equation:

\begin{thm}
\label{theo2}
Let $\Xi^{(0)} \equiv \left\{ \xi_0^{(1)}(x), \dots, \xi_0^{(N)}(x)\right\}$ be a set of arbitrary functions defined on the interval $(-L,L)$. 
We assume that these functions have a nonzero overlap with the first $N$ solutions of the inhomogeneous Helmholtz equation.
For convenience we assume that the functions are ordered as
\beq
0 < \frac{\int_{-L}^{L}  \xi_{0}^{(1)}(x) \hat{O} \xi_{0}^{(1)}(x) dx }{\int_{-L}^{L}  \xi_{0}^{(1)}(x)^2 dx} \leq \dots
\leq \frac{\int_{-L}^{L}  \xi_{0}^{(N)}(x) \hat{O} \xi_{0}^{(N)}(x) dx }{\int_{-L}^{L}  \xi_{0}^{(N)}(x)^2 dx} \ .
\eeq

Consider the sequence of functions 
\beq
\xi^{(j)}_1(x) \equiv \sqrt{\rho(x)} \int_{-L}^x dy \left[ \kappa_1^{(j)} - \int_{-L}^y dz \sqrt{\rho(z)} \xi^{(j)}_{0}(z) \right] \ ,
\label{step1}
\eeq
with $\kappa_1^{(j)} =  \frac{1}{2L} \int_{-L}^{L} dy \int_{-L}^{y} dz \sqrt{\rho(z)} \xi_{0}^{(j)}(z)$ and $j=1,\dots, N$.

Let
\beq
\bar{\xi}_1^{(j)}(x) = \xi_1^{(j)}(x) - \sum_{k=1}^{j-1} 
\frac{\int_{-L}^{L} \bar{\xi}_1^{(k)}(x) {\xi}_1^{(j)}(x) dx}{\int_{-L}^{L} \bar{\xi}_1^{(k)}(x)^2 dx} \bar{\xi}_1^{(k)}(x)
\label{step2}
\eeq
and $\Xi^{(1)} \equiv \left\{ \bar{\xi}_1^{(1)}(x), \dots, \bar{\xi}_1^{(N)}(x)\right\}$. 
We call $\Xi^{(n)}$ the set of functions obtained after $n$ iterations. Then, for 
$n \rightarrow \infty$, $\Xi^{(n)}$ converges to the $N$ lowest eigenfunctions of eq.~(\ref{helm1d}).
\end{thm}

\begin{pf*}{Proof}
The proof is straighforward: at each iteration, the functions generated with (\ref{step1}) have their lowest energy
components inflated. The orthogonalization in eq.~(\ref{step2}) eliminates from the $j^{th}$ functions the components
corresponding to the previous $j-1$ functions, which have lower expectation values. As this procedure is iterated
only the $j^{th}$ exact mode will survive in the $j^{th}$ function. This completes the proof.
\end{pf*}

The two theorems that we have just proved allow to build {\sl iteratively} the solutions corresponding 
to the lowest modes of the inhomogeneous string. A different strategy to calculate the ground state
of the string consists of using eq.~(\ref{iter}) in a non iterative way applying the Variational Theorem:
if we let $\xi_0(x)$ be an arbitrary function, depending on one or more parameters $u_i$, then
one may determine these parameters minimizing the Rayleigh quotient corresponding to $\xi_1(x)$, i.e.
the function obtained with eq.~(\ref{iter}) applied to $\xi_0(x)$.
Alternatively one could minimize the Rayleigh quotient corresponding to $\xi_0(x)$, but this is clearly
less efficient, since $\xi_1(x)$ is always closer to the fundamental mode because of Theorem \ref{theo1}.

We may also formulate a non-iterative procedure which allows to obtain approximations for the lowest part of the spectrum of the
string. We now briefly describe this approach.

Let $\Xi^{(0)} \equiv \left\{ \xi_0^{(1)}(x), \dots, \xi_0^{(N)}(x)\right\}$ be a set of functions defined on the interval $(-L,L)$, 
and let $\Xi^{(1)} \equiv \left\{ \bar{\xi}_1^{(1)}(x), \dots, \bar{\xi}_1^{(N)}(x)\right\}$ be the set of functions
obtained after applying eq.~(\ref{iter}) to each of the functions in $\Xi^{(0)}$ and normalizing each of them:
\beq
\int_{-L}^{L} \xi_1^{(j)}(x)^2 dx = 1 \ .
\eeq 

Let $\mathbf{A}$  and $\mathbf{B}$ be the matrices whose elements are 
\beq
\mathbf{A}_{ij} &\equiv& \int_{-L}^{L} \xi_1^{(i)}(x) \left[
\frac{1}{\sqrt{\rho}} \left( -\frac{d^2}{dx}^2\right) \frac{1}{\sqrt{\rho}} \right] \xi_1^{(j)}(x) \\
\mathbf{B}_{ij} &\equiv& \int_{-L}^{L} \xi_1^{(i)}(x) \xi_1^{(j)}(x) \ .
\eeq

The eigenvalues $\lambda$ and eigenvectors $\mathbf{v}$ of the generalized eigenvalue problem 
\beq
\mathbf{A} \mathbf{v} = \lambda \mathbf{B} \mathbf{v}
\label{gep}
\eeq
are approximations to the lowest modes of the string. The advantage of this non-iterative procedure
is that the set $\Xi_0$ may be chosen arbitrarily (for example, selecting functions for which the
integrals can be done analytically) and that there is no need of orthogonalization, as in Theorem \ref{theo2}.

A yet different approach would consist of applying this procedure to a set of functions
$\bar{\Xi}_0 \equiv \left\{ \xi_0(x), \dots, \xi_N(x)\right\}$, obtained after $N$ iterations
of eq.~(\ref{iter}) on an arbitrary function $\xi_0(x)$ (we also assume that each function is normalized).
In this case $\Xi_0(x)$ defines a Krylov subspace, and it can be used to set up a generalized eigenvalue problem
as in eq.~(\ref{gep}). One expects this procedure to be more efficient that the previous, given that 
the iteration of eq.~(\ref{iter}) provides functions whose lower energy components are enhanced: however, 
from a practical point of view, the iteration of eq.~(\ref{iter}) in general generates functions which are
more and more complicated (this problem could be obviated, at least in part, by finding suitable 
analytical approximations for each function). 

The results that we have described so far allow to obtain arbitrarily precise approximations for the 
lowest modes of an inhomogeneous string.

We now enunciate a third theorem, which allows one to calculate excited states of the string without
the need of orthogonalization with respect to the lower lying modes.

\begin{thm}
\label{theo3}
Let $\eta_0(x)$ be an arbitrary function fulfilling Dirichlet boundary conditions on $[-L,L']$, with $-L < L' < L$.
The sequence of functions
\beq
\eta_n(x) \equiv \sqrt{\rho(x)} \int_{-L}^x dy \left[ \tilde{\kappa}_n - \int_{-L}^y dz \sqrt{\rho(z)} \eta_{n-1}(z) \right] \ ,
\label{iter3}
\eeq
with $\tilde{\kappa}_n =  \frac{1}{L+L'_n} \int_{-L}^{L'_n} dy \int_{-L}^{y} dz \sqrt{\rho(z)} \eta_{n-1}(z)$, converges 
to a solution of equation (\ref{helm1d}) for $n\rightarrow \infty$ if $L'$ is chosen so that
$\eta_n(L)=0$.
\end{thm}

\begin{pf*}{Proof}
The proof goes as follows: for a fixed $L'$ the sequence of functions $\eta_n(x)$ would converge to the ground state 
of the Helmholtz equation  (\ref{helm1d}) with Dirichlet boundary conditions in $x=-L$ and $x=L'$ because of Theorem
\ref{theo1}. By fixing $L'$ at each iteration so that $\eta_n(L)=0$, one obtains a function which converges to a 
solution of eq.~(\ref{helm1d}) on the whole string.
\end{pf*}

Notice that as the number $n$ of iterations is increased, the equation $\eta_n(L)=0$ acquires multiple solutions, each one
approximating a different solution of  eq.~(\ref{helm1d}).

We will apply these theorems to specific examples in Section \ref{applications}.

\section{Perturbation theory}
\label{perturbation}

While in the previous sections we made no assumption regarding the behaviour of $\rho(x)$, apart from requiring it 
to be a positive function, now we also assume it to represent a slightly inhomogeneous string . In this case 
the problem can be treated with the perturbative approach of ref. ~\cite{Amore10}. Notice that
the perturbation method developed in that paper may be regarded as a {\sl shape perturbation method}, since 
the "density" was obtained from a conformal map and therefore changes in the density corresponded to changes 
of the shape of the membrane. In this case however $\rho(x)$ is really a mass density, so that we may regard 
this specific application of perturbation theory as a genuine example of {\sl density perturbation theory}.

We briefly review here the perturbative approach described in ref.~ \cite{Amore10} and adapt it to the specific
problem that we are considering: the starting point is the symmetrized operator
\beq
\hat{O} = \frac{1}{\sqrt{\rho}} (-\Delta) \frac{1}{\sqrt{\rho}} \ ,
\label{pt1}
\eeq
which reduces to the negative laplacian for a constant $\rho$. In the present case it is understood that 
$\Delta$ is the second derivative $d^2/dx^2$.

We now express $\rho(x)$  as
\beq
\rho(x) = \rho_0 \left( 1+ \eta \sigma(x) \right)
\label{pt2}
\eeq
where $|\sigma(x)|\ll 1$ and $\rho_0 = \frac{1}{2L} \int_{-L}^{L} \rho(x) dx$ is the average mass density in $(-L,L)$.
The parameter $\eta$ has been introduced for power counting; we will  use $\eta$ to keep track of the different orders 
in powers of $\sigma(x)$ and set it to $1$ at the end.

Our operator may now be expanded in powers of $\eta$ as
\beq
\hat{O} &\approx& \frac{1}{\rho_0} \ \left[ \hat{O}_0 + \eta \hat{O}_1 + \eta^2 \hat{O}_2 + \eta^3 \hat{O}_3 + \dots \right] \ ,
\label{pt3}
\eeq
where the explicit form of the  $\hat{O}_i$ may be worked out rather easily:
\beq
\label{pt4}\hat{O}_0 &=& -\Delta \\
\label{pt5}\hat{O}_1 &=& - \frac{1}{2} \left[ \sigma (-\Delta) + (-\Delta) \sigma\right] \\
\label{pt6}\hat{O}_2 &=& \frac{1}{8} \left[ 2 \sigma (-\Delta) \sigma + 3 \sigma^2 (-\Delta) + 
3 (-\Delta) \sigma^2 \right]  \\
\hat{O}_3   &=&  - \frac{3}{16} \left[  \sigma^2 (-\Delta) \sigma +  \sigma (-\Delta) \sigma^2 \right]  \nonumber \\
\label{pt7} &-& \frac{5}{16} \left[  \sigma^3 (-\Delta)  +   (-\Delta) \sigma^3 \right]   \\
\dots && \nonumber
\eeq

This problem may be treated within the standard Rayleigh-Schr\"odinger perturbation theory (RSPT), keeping
in mind that the operator $\hat{O}$ contains the perturbation to all orders in $\eta$ (typical applications
of RSPT see the perturbation term in the Hamiltonian to be of order one in the coupling constant).

Working up to third order one finds
\beq
\label{pt8}\rho_0 \ E_n^{(0)} &=& \epsilon_n \\
\label{pt9}\rho_0 \ E_n^{(1)} &=& \langle n | \hat{O}_1| n \rangle \\
\label{pt10}\rho_0 \ E_n^{(2)} &=& \langle n | \hat{O}_2| n \rangle + 
\sum_{k\neq n} \frac{|\langle n | \hat{O}_1 | k \rangle|^2}{\epsilon_n -\epsilon_k} \\
\rho_0 \ E_n^{(3)} &=& \langle n | \hat{O}_3| n \rangle + 2
\sum_{k\neq n} \frac{\langle n | \hat{O}_2 | k \rangle \langle k | \hat{O}_1 | n \rangle}{\epsilon_n -\epsilon_k} 
\nonumber \\
&+& \sum_{k\neq n} \sum_{m \neq n} 
\frac{\langle n | \hat{O}_1 | m \rangle \langle m | \hat{O}_1 | k \rangle \langle k | 
\hat{O}_1 | n \rangle}{(\epsilon_n -\epsilon_k) (\epsilon_n -\epsilon_m)} \nonumber \\
\label{pt11}&-& \langle n | \hat{O}_1 | n \rangle 
\sum_{k\neq n} \frac{ \langle n | \hat{O}_1 | k \rangle^2}{(\epsilon_n -\epsilon_k)^2} \ ,
\eeq
where $\epsilon_n = \frac{n^2\pi^2}{4L^2}$  and $|n\rangle$ are the eigenvalues and eigenstates of $-d^2/dx^2$.

As shown in \cite{Amore10} one may work out the explicit form of these equations and obtain the formulas:
\beq
\label{pt11b}\rho_0 \ E_n^{(0)} &=& \epsilon_n \\
\label{pt12}\rho_0 \ E_n^{(1)} &=& - \epsilon_n \langle n | \sigma | n \rangle  \\
\label{pt13}\rho_0 \ E_n^{(2)} &=& \epsilon_n \langle n | \sigma | n \rangle^2 
+ \epsilon_n^2 \sum_{k \neq n} \frac{\langle n | \sigma | k \rangle^2 }{\epsilon_n-\epsilon_k} \\
\rho_0 \ E_n^{(3)} &=& - \epsilon_n \langle n | \sigma | n \rangle^3 + \epsilon_n^3 \langle n | \sigma | n \rangle
\sum_{k \neq n}  \frac{\langle n | \sigma | k \rangle^2}{\omega_{nk}^2} \nonumber \\
&-& 3 \epsilon_n^2  \langle n | \sigma | n \rangle \sum_{k \neq n}  
\frac{\langle n | \sigma | k \rangle^2}{\omega_{nk}} \nonumber \\
\label{pt14}&-& \epsilon_n^3 \sum_{k\neq n} \sum_{m \neq n} 
\frac{\langle n | \sigma | k \rangle \langle k | \sigma | m \rangle 
\langle m | \sigma | n \rangle }{\omega_{nk}\omega_{nm}}  \ .
\eeq

Unlike in the cases studied in ref.~\cite{Amore10} these formulas apply to all the spectrum, which in one 
dimension is nondegenerate.

If one neglects in these formulas the terms mixing different states, one may see that the remaining terms correspond to the 
terms of a geometric series and obtain the approximation (see \cite{Amore10})
\beq
E_n \approx \frac{\epsilon_n}{\langle n | \rho_0 (1+\sigma(x)) |n\rangle }  
= \frac{\epsilon_n}{\langle n | \rho(x) |n\rangle }  \ .
\eeq

This resummation was used in \cite{Amore10b} to rederive Weyl's law for two dimensional membranes and to
generalize the law to two dimensional membranes of arbitrary density and shape (and in higher dimensions
to $d$-cubes of arbirtrary density). We do not have a general proof that the terms mixing different states
can always be neglected, although the fact that Weyl's law is obtained in this limit may support this claim.
In the case of the small deformations of a square drum, we have proved in  \cite{Amore10} that the contributions
which mix different states at second order in perturbation theory are subleading with respect to the 
"diagonal" contributions.

We will now show that in one dimension the Weyl's density law is modified.

To start with we notice that for $n \rightarrow \infty$
\beq
\langle n | \sigma(x) |n \rangle \approx \frac{1}{2L} \int_{-L}^{L}  \sigma(x) \ dx \ .
\eeq
Notice that in two dimensions, in the cases in which $\sigma(x,y)$ is related to a conformal density, 
one recovers in this limit the area of the membrane (see \cite{Amore10}), from which Weyl's law follows.

Let us now focus of the second contribution in eq.(\ref{pt13}), which contains matrix elements of $\sigma$ 
between different states:
\beq
\epsilon_n \sum_{k \neq n} \frac{\langle n | \sigma | k \rangle^2 }{\epsilon_n-\epsilon_k} \ .
\eeq
We are interested in the behaviour of this term for $n\rightarrow \infty$.

Let us consider an arbitrary $\sigma(x)$ and write
\beq
\langle n | \sigma | k \rangle \approx \frac{1}{2L} \int_{-L}^{+L} \cos \left[ \frac{(n-k) \pi}{2L} (x+L)\right] \sigma(x) dx
\eeq
which holds for $n\rightarrow \infty$.

Therefore we have
\beq
\epsilon_n \sum_{k \neq n} \frac{\langle n | \sigma | k \rangle^2 }{\epsilon_n-\epsilon_k} \approx
\int_{-L}^{+L} dx \int_{-L}^{+L} dy \ \sigma(x) \Delta(x,y)\sigma(y) 
\label{perturb2}
\eeq
where
\beq
\Delta(x,y) \equiv \frac{1}{4L^2} \sum_{k\neq n} \frac{n^2}{n^2-k^2} 
\cos \left[ \frac{(n-k) \pi}{2L} (x+L)\right] \cos \left[ \frac{(n-k) \pi}{2L} (x+L)\right] \ .
\eeq

It is convenient to decompose $\sigma(x)$ in a even and odd functions and therefore write
\beq
\epsilon_n \sum_{k \neq n} \frac{\langle n | \sigma | k \rangle^2 }{\epsilon_n-\epsilon_k} \approx
\int_{-L}^{+L} dx \int_{-L}^{+L} dy \left[ \sigma_e(x) \Delta^{(e)}(x,y)\sigma_e(y)  + 
\sigma_o(x) \Delta^{(o)}(x,y)\sigma_o(y) \right]
\eeq
where the sum in $\Delta^{(e)}(x,y)$ ($\Delta^{(o)}(x,y)$ )
contains the values of $k \neq n$, with $|k-n|$ even (odd).

Let us first consider the function corresponding to odd indices:
\beq
 \Delta^{(o)}(x,y) &\approx& \sum_{j=0}^{\infty} \frac{\cos \left(\frac{\pi  (2 j+1) (L+x)}{2 L}\right) 
\cos \left(\frac{\pi  (2 j+1) (L+y)}{2 L}\right)}{8 L^2} \nonumber \\
&=& \frac{1}{8L} \sum_{j=0}^{\infty} \phi_j(x) \phi_j(y)
\eeq
where $\phi_j(x)$ are the odd functions of the orthonormal basis fulfilling Neumann boundary conditions
on the interval $[-L,L]$ (see eq.~(20) of \cite{Amore09}). Keeping in mind that $\Delta^{(o)}(x,y)$ only
acts on odd functions, we may use the completeness of this set to make the following identification:
\beq
\Delta^{(o)}(x,y) &\approx& \frac{\delta(x-y)}{8L} \ .
\eeq 

We may now come to the function corresponding to even indices:
\beq
 \Delta^{(e)}(x,y) &\approx& \sum_{j=1}^{\infty} \frac{\cos \left(\frac{\pi j (L+x)}{L}\right) 
\cos \left(\frac{\pi j (L+y)}{L}\right)}{8 L^2} \nonumber \\
&=& \frac{1}{8L} \sum_{j=1}^{\infty} \psi_j(x) \psi_j(y)
\eeq
where $\psi_j(x)$ are the even functions  of the orthonormal basis fulfilling Neumann boundary conditions
on the interval $[-L,L]$ (see eq.~(19) of \cite{Amore09}). To exploit the completeness relation of this basis
we must rewrite this expression including the zero mode as
\beq
 \Delta^{(e)}(x,y) &\approx& \frac{1}{8L}  \left[-\frac{1}{2L} + \sum_{j=0}^{\infty} \psi_j(x) \psi_j(y) \right] \nonumber \\
&=& -\frac{1}{16L^2} + \frac{\delta(x-y)}{8L} 
\eeq
keeping in mind that this function operates only on even functions.

We may use the results to write the asymptotic behaviour of the second order perturbative term:
\beq
\rho_0 \ \frac{E_n^{(2)}}{\epsilon_n} &=&  \langle n | \sigma | n \rangle^2 
+ \epsilon_n \sum_{k \neq n} \frac{\langle n | \sigma | k \rangle^2 }{\epsilon_n-\epsilon_k} \nonumber \\
&\approx&  \left[ \frac{1}{2L} \int_{-L}^{L} \sigma(x) dx \right]^2 + 
\frac{1}{8L} \int_{-L}^{L} \sigma^2(x) dx - \left[ \frac{1}{4L} \int_{-L}^{L} \sigma(x) dx \right]^2 \nonumber \\
&\approx& \frac{1}{4} \langle  \sigma^2\rangle  + \frac{3}{4} \langle \sigma \rangle^2 \ .
\label{pert2}
\eeq

Notice that this behaviour is compatible with the form of Weyl's law for a string 
of variable density $\rho(x) = \rho_0 (1+\sigma(x))$ of the form
\beq
E_n \approx \frac{\epsilon_n}{\langle \sqrt{\rho(x)} \rangle^2}
\label{weyl1d}
\eeq
where $\langle \sqrt{\rho} \rangle \equiv \frac{1}{2L} \int_{-L}^L \sqrt{\rho(x)} dx $.
As a matter of fact we may expand eq.~(\ref{weyl1d}) in powers of $\sigma$ and obtain:
\beq
E_n &\approx& \frac{\epsilon_n}{\rho_0} \ \left[ 1 - \langle \sigma \rangle + \left(\frac{1}{4} \langle \sigma^2\rangle
+ \frac{3}{4} \langle \sigma\rangle^2 \right)+ \dots \right] \ ,
\eeq
which to second order reproduces the perturbative result that we have just discussed. 
Actually, one may derive this result using the WKB method, as done by Bender and Orszag in 
their classical book, \cite{BO78}, eq.(10.1.31) pag.490. 
For the case of a string of density which varies linearly with the position, Crawford \cite{Crawford87} has compared the WKB
formula with the exact results obtained by Fulcher for this particular problem, see \cite{Fulcher85}, showing that it is 
very accurate.

In a similar way one may also obtain the perturbative corrections to the states of the string: in this case,
working to first order one obtains
\beq
|\Psi_n^{(0)}\rangle &=& | n \rangle \\
|\Psi_n^{(1)}\rangle &=& -\frac{1}{2} \sum_{k\neq n} \langle k | \sigma | n\rangle  
\frac{\epsilon_n^{(0)}+\epsilon_k^{(0)}}{\epsilon_n^{(0)}-\epsilon_k^{(0)}} | k \rangle \nonumber \\
&=& - \epsilon_n^{(0)} \ \sum_{k\neq n}  \frac{\langle k | \sigma | n\rangle}{\epsilon_n^{(0)}-\epsilon_k^{(0)}} | k \rangle +
\frac{1}{2} \left[ \sigma(x) - \langle n | \sigma | n \rangle \right] |k\rangle 
\eeq
where $|n\rangle$ are the unperturbed states.

A general expression for the solutions of a string of variable density obtained with the WKB method is 
also given in eq.~(10.1.33) of ref.~\cite{BO78}:
\beq
\psi_n(x) \propto \left(\int_{-L}^{L} \frac{\sqrt{\rho(t)}}{2} dt \right)^{-1/2} \ \rho(x)^{-1/4} \ 
\sin \left[ n\pi \frac{\int_{-L}^{x} \sqrt{\rho(t)} dt}{\int_{-L}^{L} \sqrt{\rho(t)} dt}
\right] \ ,
\label{wkbwf}
\eeq
which is valid for $n\rightarrow \infty$.

\section{Applications}
\label{applications}

In this Section we consider two  examples of strings of variable density and apply to these problems
the numerical and analytical results obtained in the previous sections. We adopt the convention of working
with strings of unit length, centered in the origin.

\subsection{"Borg string"}

The first example that we consider has been originally discussed by Borg in \cite{Borg46} and corresponds
to a density
\beq
\rho(x) = \frac{(1+\alpha)^2}{(1+\alpha (x+1/2))^4}  \ ,
\eeq
where $\alpha>-1$ is a free parameter and $x \in (-1/2,1/2)$.
This string is isospectral to a string of uniform density $\rho(x) =1$, and its solutions
are known explicitly.

The conditions for the isospectrality of two strings of different density have been recently discussed by
Gottlieb in ref.~\cite{Gottlieb02}, where the "Borg string" is one the examples considered in that paper.

Gottlieb shows that the nonuniform Helmholtz's equations  
\beq
- \frac{d^2}{d\xi^2} \psi(\xi) &=& E \rho_1(\xi) \psi(\xi) \\
- \frac{d^2}{dx^2} \phi(x) &=& E \rho_2(x) \phi(x) \ ,
\eeq
are isospectral if $\xi(x) = \frac{ (x+1/2) (1+\alpha)}{1+\alpha (x+1/2)} - \frac{1}{2}$ 
(with $\alpha>-1$ and $x,\xi \in (-1/2,1/2)$) and
\beq
\rho_2(x) = \left(\frac{d\xi}{dx}\right)^2 \rho_1(\xi(x)) .
\eeq

The reader should notice that this relation is fully consistent with the asymptotic law in eq.~(\ref{weyl1d}), 
since it implies
\beq
\int_{-1/2}^{1/2} \sqrt{\rho_2(x)} dx  = \int_{-1/2}^{1/2} \sqrt{\rho_1(\xi)} d\xi .
\eeq

In the case of a uniform string, the rational transformation considered by Gottlieb, provides 
precisely the density of the "Borg string".

For this problem the asymptotic WKB method provides the exact energies and solutions;
as a matter of fact, in the case of the energy, eq.~(\ref{weyl1d}), reduces to $E_n  = n^2\pi^2$, 
since $\langle \sqrt{\rho} \rangle = 1$; in the case of the wave functions, the WKB formula of eq.~(\ref{wkbwf})
reduces to 
\beq
\psi_n(x) = \frac{(2 \alpha  x+\alpha +2)}{\sqrt{2 (\alpha +1)}} \ 
\sin \left(\frac{\pi  (\alpha +1) n (2 x+1)}{2 \alpha  x+\alpha +2}\right) \ ,
\label{borg_exact}
\eeq
which agrees with eq.~(4.5) of Gottlieb 2002 apart from a normalization factor.

We will now use the Borg string to test the numerical and analytical methods of our paper.

\begin{figure}
\begin{center}
\bigskip\bigskip\bigskip
\includegraphics[width=6.5cm]{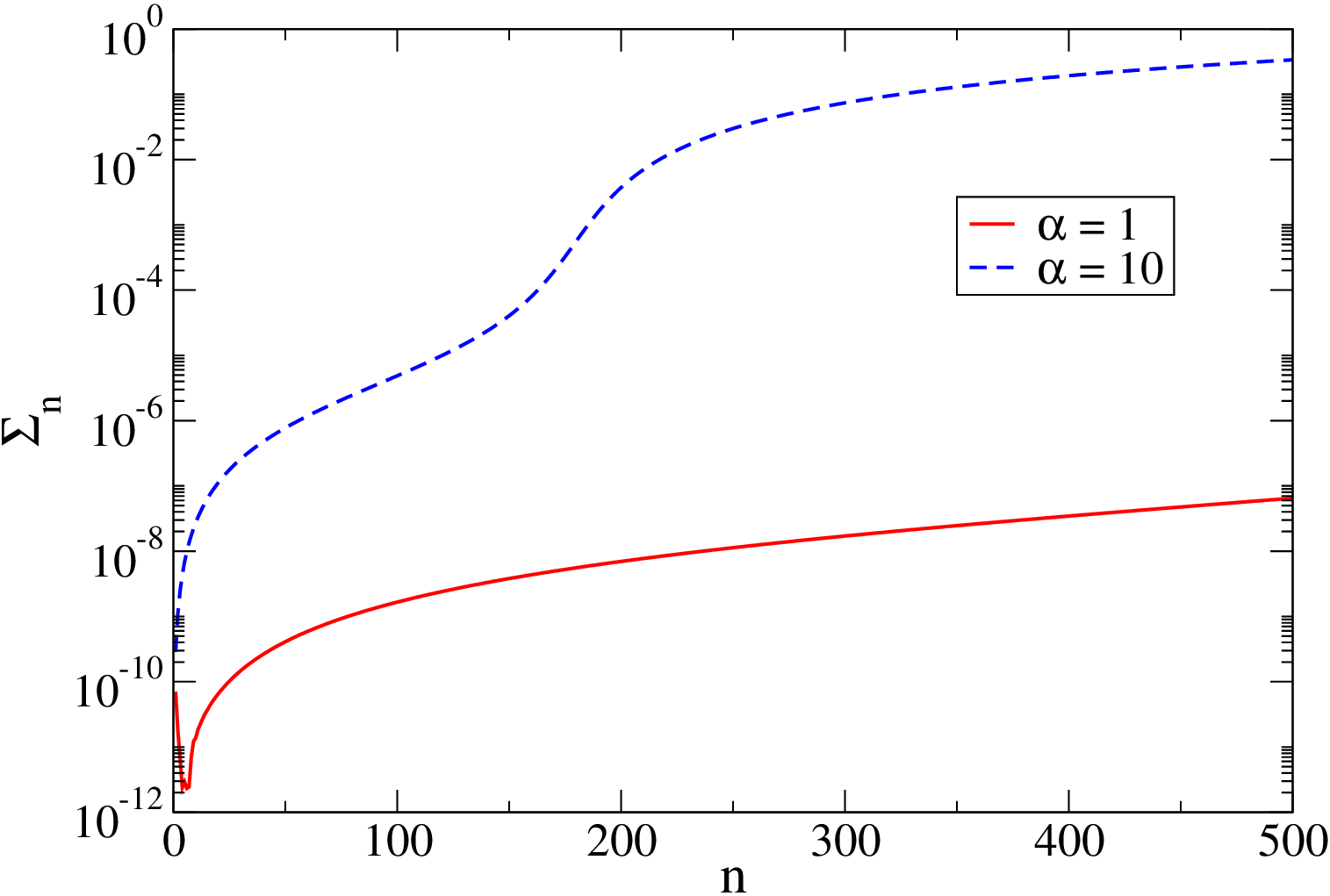}
\includegraphics[width=6.5cm]{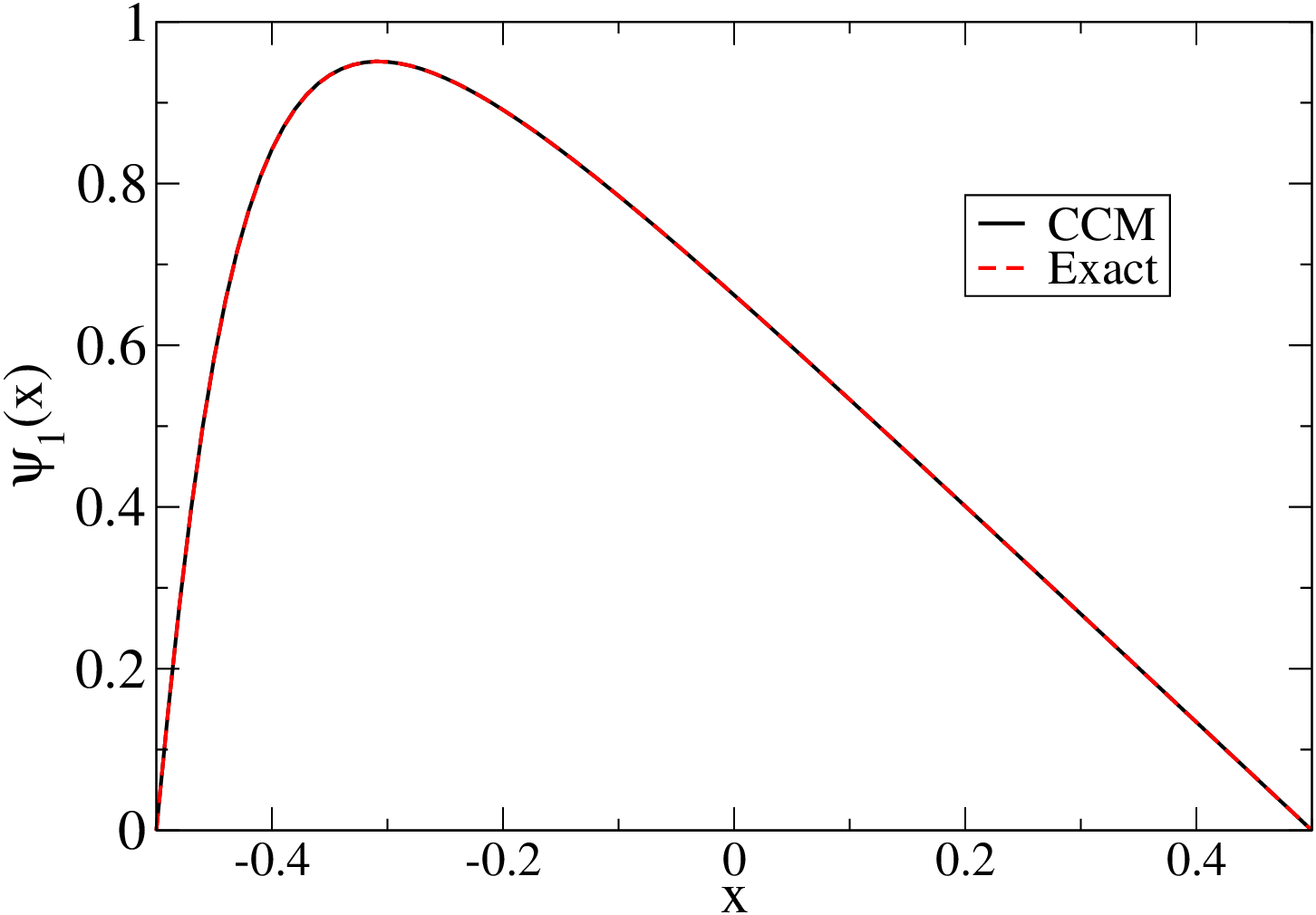}
\caption{Left plot: Error over the energies of the Borg string, $\Sigma_n = E_n^{CCM}/E_n^{exact}-1$, for $\alpha=1$ (solid line)
and for $\alpha=10$ (dashed line), calculated with a grid corresponding to $N=2000$.
Right plot:Solution for the fundamental mode of the Borg string for $\alpha=10$. The dashed line is the exact result
$\psi_1(x) = 2 \sqrt{\frac{2}{11}} (5 x+3) \sin \left(\frac{11 \pi  (2 x+1)}{4 (5 x+3)}\right)$. The solid line is 
the numerical result obtained with the collocation method using a grid with $N=2000$.}
\label{Fig_1}
\end{center}
\end{figure}

We first focus on the application of the collocation method of Section \ref{collocation}:
in the left plot of Figure \ref{Fig_1} we display the error $\Sigma_n = E_n^{CCM}/E_n^{exact}-1$, 
for $\alpha=1$ (solid line) and for $\alpha=10$ (dashed line), where $E_n^{CCM}$ are the energies 
calculated with the collocation method of section \ref{collocation} using  a grid 
corresponding to $N=2000$. Notice that for $\alpha = 10$ we are considering a strongly inhomogeneous string and
the precision of the numerical results is clearly smaller, compared to the precision of the results
corresponding to $\alpha = 1$. An intuitive justification of this phenomenon is the following: as $\alpha$ 
grows, the density becomes strongly peaked around on of the ends of the string; in this case the excited modes of the string
tend to oscillate more rapidly in this region and more slowly in the lower density region, compared to the corresponding 
modes of a uniform string. In the collocation approach, the number of grid points determines the maximal resolution 
which can be achieved in the calculation: for this reason the numerical precision obtained for the inhomogeneous modes, 
which need better resolution, is necessarily poorer.

In the right plot of Figure \ref{Fig_1} we compare the exact and numerical solution for the  fundamental mode 
of the Borg string for $\alpha = 10$. The two curves are not distinguishable: the integrated error for the solutions is
$\Xi \equiv \int_{-L}^{L} \left| \psi_1^{CCM}(x) - \psi_1^{exact}(x) \right| \approx 3.36 \times 10^{-10}$,
which is consistent with the error over the energy for the fundamental mode in the left plot of the figure.

We would like to draw the attention of the reader on the flexibility of the collocation method: the calculation
for strings of different density can be made quite efficiently specifying the density profile and evaluating it
on an uniform grid. The construction of the diagonal density matrix is extremely fast and it takes typically
a fraction of a second on an average desktop computer; the construction of the non-diagonal matrix for the laplacian
is more involved, although it is general and it can be  calculated once and stored. 

We now come to illustrate the application of Theorem \ref{theo1}. We pick the function $\xi_0(x) = 1$ and apply to
it eq.~(\ref{iter}) finding
\beq
\xi_1(x) = -\frac{2 (\alpha +1)^2 ((2 x+1) \log (\alpha +1)-2 \log (2 \alpha  x+\alpha +2)+\log (4))}{\alpha ^2 (2 \alpha  x+\alpha +2)^2} 
\ ,
\eeq
which now vanishes at the ends of the string, $\xi_1(\pm L) = 0$ (notice that this function is not normilized).
We may easily iterate eq.(\ref{iter}) and calculate explicitly the next functions $\xi_2(x)$, $\xi_3(x)$, \dots. 
No numerical approximation is made in this process so that one does not need to worry about round-off errors.
In this way we have obtained the expressions up to $\xi_5(x)$, although we do not report them here given their lengthy expressions.

One may now consider the normalized function corresponding to the $j^{th}$ iteration and decompose it in terms of the
exact solutions, which form an orthonormal and complete basis:
\beq
\bar{\xi}_j(x) =  \sqrt{\rho(x)} \ \sum_{k=1}^\infty \kappa_k^{(j)} \psi_{k}(x)  \ .
\eeq
where $\bar{\xi}_{j}(x) \equiv \xi_j(x)/\sqrt{\int_{-L}^L \xi_j^2(x)dx}$. The coefficients of this expansion
are
\beq
\kappa_k^{(j)} = \int_{-L}^{L} \sqrt{\rho(x)} \ \bar{\xi}_j(x) \psi_k(x) dx \ ,
\eeq
with $\sum_{k=1}^\infty \kappa_j^2 = 1$ (Parseval relation).

\begin{table}
\begin{center}
\begin{tabular}{|c||c|c|c|c|c|c||}
	\hline
$j$	&  $0$ & $1$    &   $2$  & $3$  & $4$ & $5$ \\
	\hline \hline
$\kappa_1^{j}$  & 0.59542214 & 0.98958778 & 0.99947247 & 0.99996837  & 0.99999804 & 0.99999988 \\
	\hline
\end{tabular}
\caption{First coefficient of the expansion of the functions $\bar{\xi}_j(x)$ corresponding
to different iterations of eq.(\ref{iter}) for the Borg string.}
\label{table-1}
\end{center}
\end{table}

\begin{figure}
\begin{center}
\bigskip\bigskip\bigskip
\includegraphics[width=9cm]{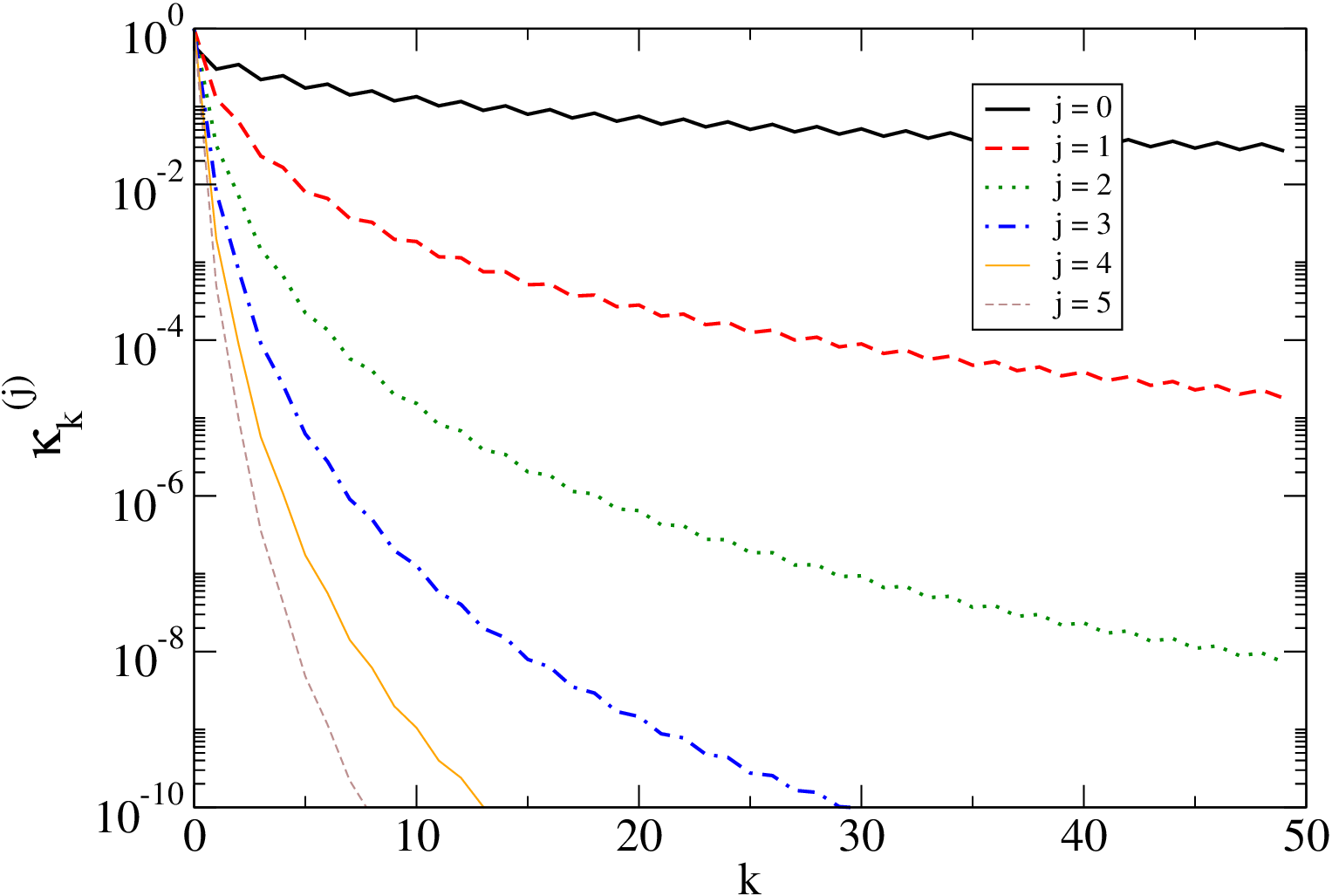}
\caption{Coefficients $\kappa_k^{(j)}$ of the expansion of the $j^{th}$ iteration $\bar{\xi}_{j}(x)$ in terms of the
exact solutions of the Borg string for $\alpha = 10$.}
\label{Fig_2}
\end{center}
\end{figure}

Figure \ref{Fig_2} and Table \ref{table-1} illustrate the convergence of the $\xi_j(x)$ towards the fundamental
mode of the string: at each iteration the component of the function obtained from eq.~(\ref{iter}) 
corresponding to the fundamental mode is inflated with respect to remaining components and the approximate solution 
gets closer to  the exact solution.

We may calculate the expectation value of $\hat{O} \equiv \frac{1}{\sqrt{\rho}} \left[ -\frac{d^2}{dx^2}\right] \frac{1}{\sqrt{\rho}}$
in each of the states corresponding to different iterations. For instance:
\beq
E_0^{(1)}(\alpha) \equiv \int_{-L}^{L} \bar{\xi}_1(x) \hat{O} \bar{\xi}_1(x) dx = \frac{n_1(\alpha) + n_2(\alpha) \log^2(1+\alpha)}{d_1(\alpha)
+ d_2(\alpha) \log(1+\alpha) + d_3(\alpha) \log^2(1+\alpha)} \ ,
\eeq
where
\beq
n_1(\alpha) &=& 108 \alpha^4 \ \  , \ \ 
n_2(\alpha) = 108 \alpha^2 (1+\alpha) \\
d_1(\alpha) &=& 8 \alpha ^2 \left(\alpha ^2+3 \alpha +3\right) \ \  , \ \ 
d_2(\alpha) = -30 \alpha  \left(\alpha ^2+3 \alpha +2\right) \\
d_3(\alpha) &=& 36 (1+\alpha)^2 \ .
\eeq

It is interesting to notice that this expression has a finite limit both for $\alpha \rightarrow -1$ and for $\alpha\rightarrow \infty$,
which correspond to the worse case scenarios for this problem. Notice that the two limit are clearly the same since the density is
invariant under the transformations $x \rightarrow -x$ and $\alpha \rightarrow - \alpha/(1+\alpha)$: therefore the Helmholtz equation
(\ref{helm1d}) is also invariant under these transformations. The reader may explicitly check that the exact solutions of 
eq.~(\ref{borg_exact}) are unchanged under the transformations above.

As a matter of fact we have
\beq
\lim_{\alpha\rightarrow -1} E_0^{(1)}(\alpha) = \lim_{\alpha\rightarrow \infty} E_0^{(1)}(\alpha) = \frac{27}{2} \ ,
\eeq
which should be compared with the exact result $E_0^{exact}= \pi^2$. The first iteration therefore provides an estimate
for the energy of the fundamental mode with an error which is at most of about $37 \%$ (for the case $\alpha=10$ studied
before the error is of about $10\%$).

The expressions corresponding to higher iterations can also be calculated explicitly, although we do not report them here;
in Table \ref{table-2} we display the energies corresponding to $\alpha = -1$ and $\alpha=\infty$ for the different
iterations and the corresponding error (in $\%$) calculated with respect to the exact result.

\begin{table}
\begin{center}
\begin{tabular}{|c||c|c|c|c|c||}
	\hline
$j$	& $1$    &   $2$  & $3$  & $4$ & $5$ \\
	\hline \hline
$\left. E_0^{(j)} \right|_{\alpha=-1,\infty}$
  & $\frac{27}{2}$ & $\frac{75117}{7550}$ & $\frac{9800153}{992524}$ &  $\frac{88566604437}{8973430934}$  & 
$\frac{1113581185772498961}{112829174318220286}$\\
Error ($\%$) & $36.7836$ & $0.807197$ & $0.0442409$ & $0.00269973$ & $0.000167958$ \\
	\hline
\end{tabular}
\caption{Expectation values of the operator $\hat{O}$ in the states corresponding to different iterations and corresponding
error.}
\label{table-2}
\end{center}
\end{table}

\begin{figure}
\begin{center}
\bigskip\bigskip\bigskip
\includegraphics[width=9cm]{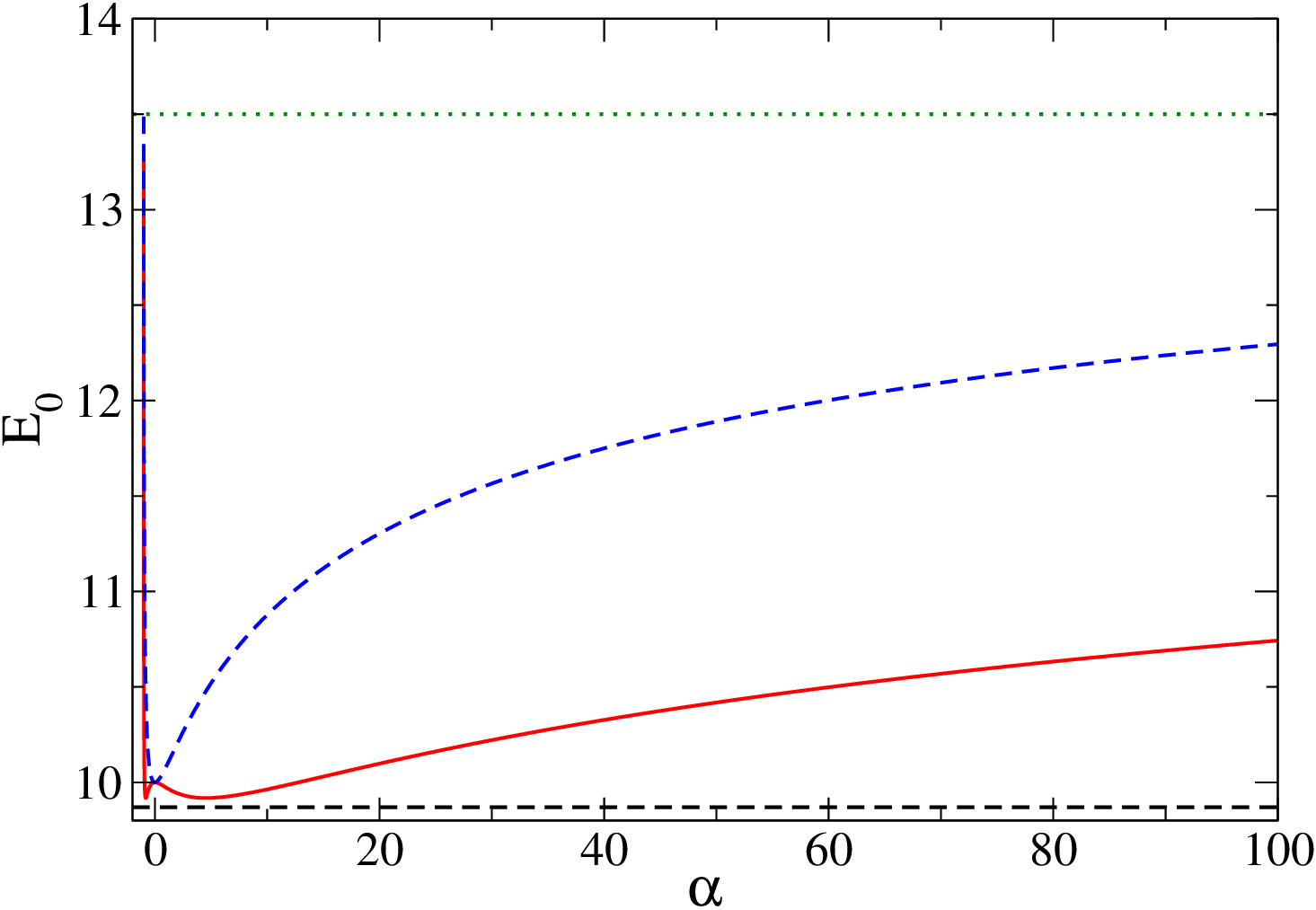}
\caption{Variational energy for the ground state obtained using $\xi_0(x) = 1 + \upsilon x$ (solid curve)  
compared with the exact result (horizontal solid line) 
and with the result obtained setting to zero the variational parameter; the upper horizontal line is the upper
bound $27/2$.}
\label{Fig_3}
\end{center}
\end{figure}

A different strategy for calculating the fundamental mode of a nonuniform string is to use the variational theorem.
We pick an arbitrary function $\xi_0(x) = 1 + \upsilon x$, where $\upsilon$ is a variational parameter which 
can be fixed by minimizing the expectation value of $\hat{O}$ in the state $\xi_1(x)$ obtained applying eq.~(\ref{iter})
to $\xi_0(x)$. In Fig.~\ref{Fig_3} we compare this variational energy with the energy obtained setting
the parameter $\upsilon=0$, which corresponds to the case studied earlier. The variational energy leads to a
consistent gain in precision for moderate values of $\alpha$, although for $\alpha \rightarrow -1$ and $\alpha\rightarrow \infty$
the two approximations provide the same limit, meaning that the term $\upsilon x$ in $\xi_0(x)$ is inessential in this regime.
Clearly, this problem may be easily obviated by using different variational ansatz with the correct asymptotic behaviour and/or
by using ansatzes with multiple variational parameters (possibly, simple enough so that the integrals in eq.(\ref{iter}) can
be done explicitly).

\begin{figure}
\begin{center}
\bigskip\bigskip\bigskip
\includegraphics[width=9cm]{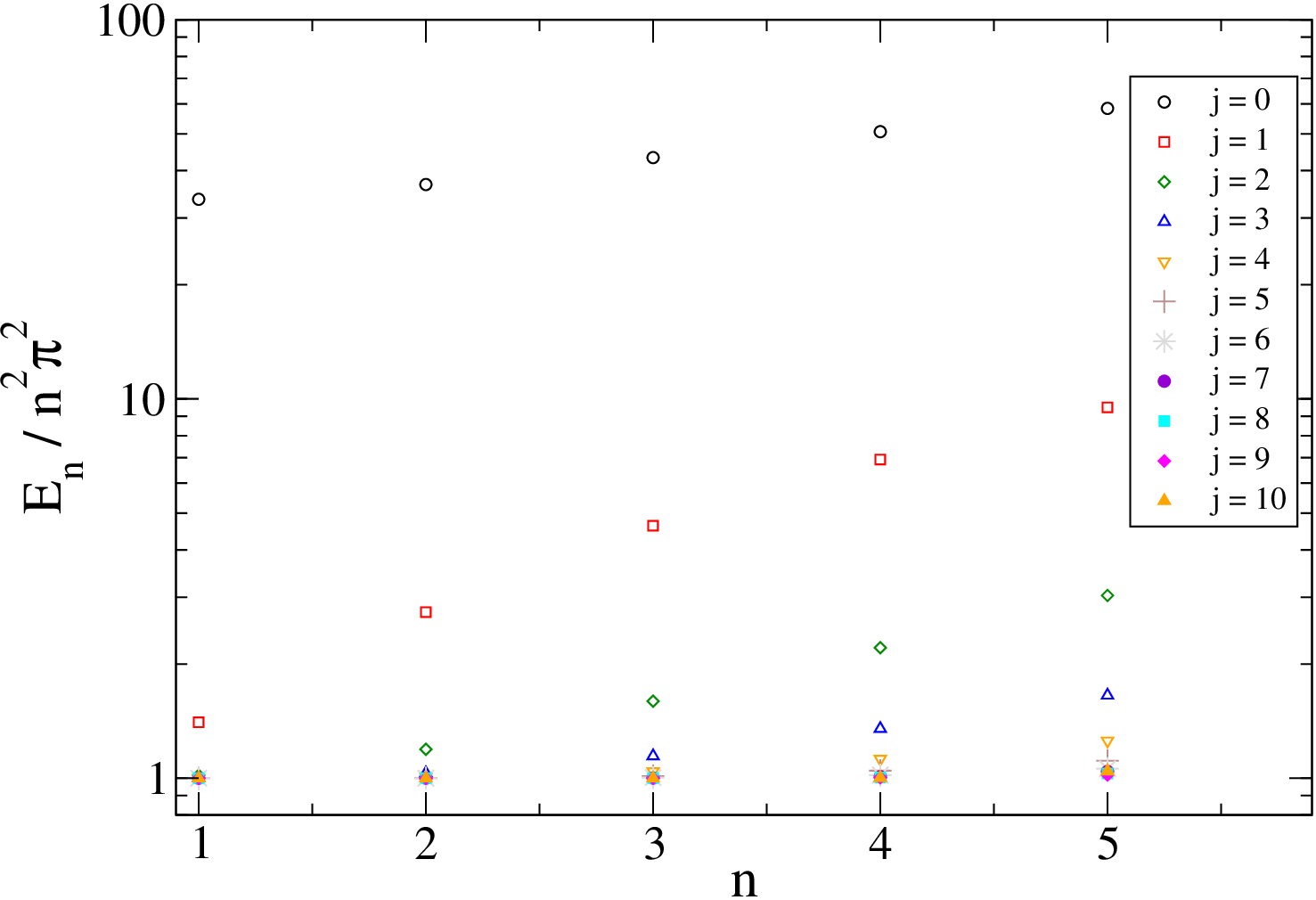}
\caption{Application of Theorem \ref{theo2} to the Borg string with $\alpha = 10$. The sets represent
the ratio $E_{n}/n^2\pi^2$ for the first five modes ($n=1,\dots, 5$) corresponding to the different iterations. }
\label{Fig_4}
\end{center}
\end{figure}

In Fig.~\ref{Fig_4} we illustrate Theorem \ref{theo2} on the first $5$ states of the Borg string, for $\alpha=10$.
We have used as starting functions the functions
\beq
\xi_0^{n}(x) = \frac{3 \sqrt{2} \left(1-4 x^2\right) C_{n+1}^{\left(\frac{5}{2}\right)}(2 x)}{\sqrt{\frac{\Gamma (n+6)}{\left(n+\frac{7}{2}\right)
   (n+1)!}}} \ ,
\eeq
where $C_{n+1}^{\left(\frac{5}{2}\right)}(2 x)$ are the Gegenbauer polynomials (we have chosen these functions because they are
simple enough to perform explicit calculations). These functions are not a particularly good ansatz, as
we can appreciate by looking at the open circles in the Figure, which show that each of these functions has an average energy which 
is far higher than the corresponding exact values. At each iteration however the ratio decreases and tends monotonically to $1$
for all the states. From a practical point of view this Theorem can be used to obtain approximations for the few lowest energy modes,
given that each new state needs to be orthogonalized with respect to the previous one, thus leading to a rapid increase in the number
of operations needed.

\subsection{Parabolic density}

We consider now the case of a string with parabolic density
\beq
\rho(x) = ( 1 + \alpha x)^2
\eeq
where $|\alpha |\leq 2$ implies that $\rho(x)>0$ for $x \in (-1/2,1/2)$. The reason for studying this string is
practical, since it allows a simple implementation of Theorem \ref{theo3}.

We choose $\eta_0(x) = (x+1/2) (L'-x)$ and apply eq.(\ref{iter3}) to calculate the higher iterations, leaving $L'$ 
as a shooting parameter. For example, after one iteration we find:
\beq
\eta_1(x) = \eta_0(x)  \left[  a_0 + a_1 x + a_2 x^2 + a_3 x^3 + a_4 x^4 \right]
\eeq
where
\beq
a_0 &=& \frac{1}{240} \left(\alpha  \left(8 {L'}^3+6 {L'}^2-3 {L'}-1\right)+5 \left(4 {L'}^2+6 {L'}+1\right)\right) \nonumber \\
a_1 &=& \frac{1}{240} \left(\alpha ^2 \left(8 {L'}^3+6 {L'}^2-3 {L'}-1\right)+\alpha  \left(28 {L'}^2+36
   {L'}+7\right)+20 {L'}-10\right) \nonumber  \\
a_2 &=& \frac{1}{120} \left(\alpha ^2 \left(4 {L'}^2+3 {L'}+1\right)+7 \alpha  (2 {L'}-1)-10\right) \nonumber  \\
a_3 &=& \frac{1}{60} \alpha  (\alpha  (2 {L'}-1)-8) \nonumber \\
a_4  &=& -\frac{\alpha ^2}{20} \nonumber \ .
\eeq

The condition $\eta_1(L) = 0$, provides the quartic equation 
\beq
(\alpha +2) (2 {L'}-1) (\alpha  (4 {L'} ({L'}+1) (4 {L'}+1)-5)+10 (4 {L'} ({L'}+2)-1)) = 0 \ ,
\label{shoot}
\eeq
which has a fixed solution, corresponding to $L'=1/2$ and three remaining solutions which depend on $\alpha$.
Although these last solutions can be calculated explicitly, their expressions are lengthy and not particularly enlightening: 
we prefer to give approximate expressions around $\alpha=0$:
\beq
L'_a &=& \frac{1}{2} \\
L'_b &\approx& \frac{1}{10} \left(\sqrt{5}-2\right) (2 \alpha +5) + O\left[\alpha^2\right] \ , .\nonumber \\
L'_c &\approx& \frac{4 \alpha }{5}-\frac{5}{2 \alpha }+\frac{3}{4} + O\left[\alpha^2\right] \ , \nonumber \\
L'_d &\approx& -\frac{1}{10} \left(2+\sqrt{5}\right) (2 \alpha +5)+ O\left[\alpha^2\right] \ ,\nonumber \ .
\eeq

\begin{figure}
\begin{center}
\bigskip\bigskip\bigskip
\includegraphics[width=9cm]{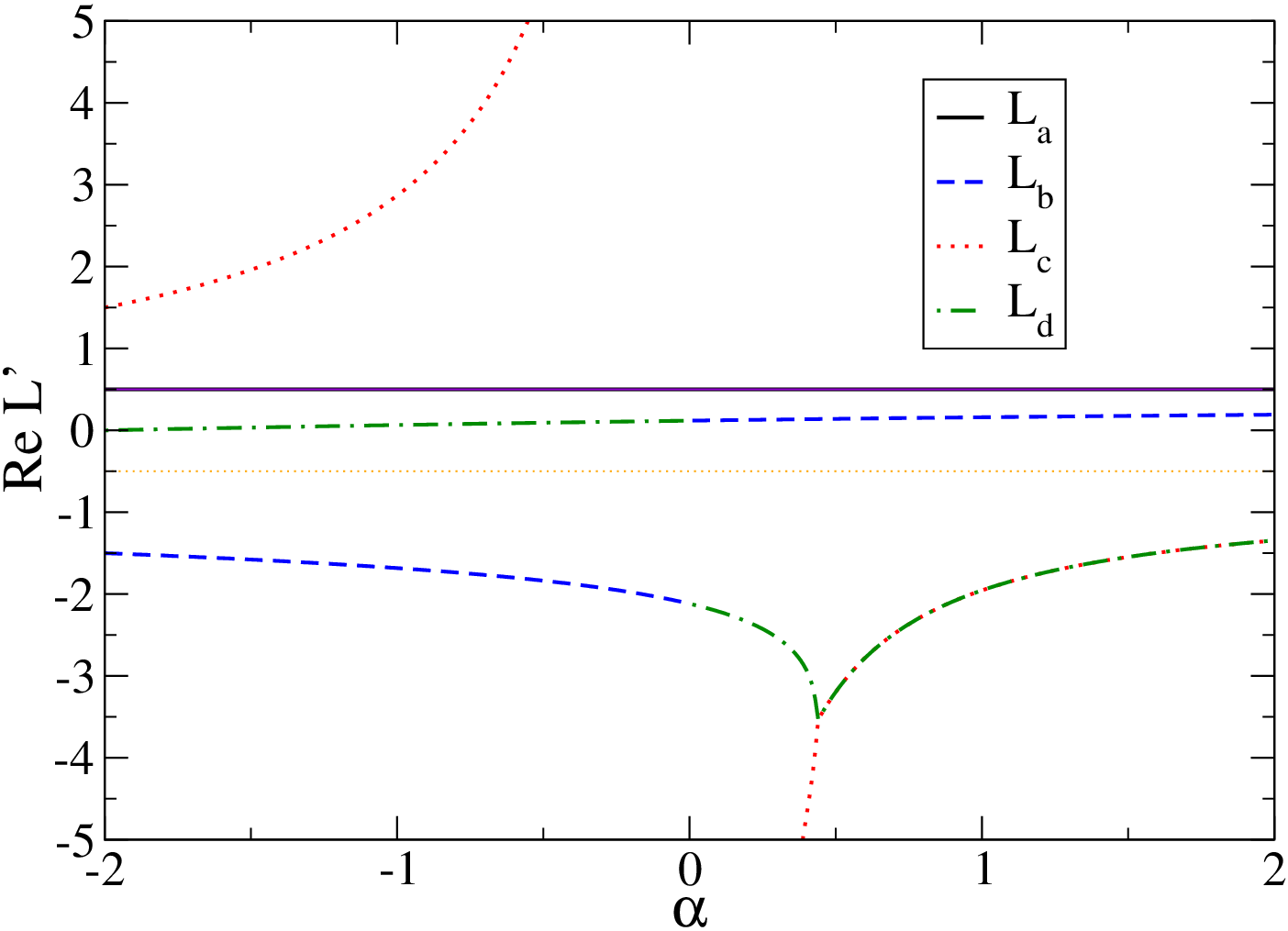}
\caption{Real part of the solutions to eq.~(\ref{shoot}).}
\label{Fig_5}
\end{center}
\end{figure}

In Fig.~\ref{Fig_5} we draw the four solutions (just the real part). In order to be acceptable, the solutions must be real and
$-L \leq L'\leq L$, i.e. they must fall in the region delimited by the two horizontal lines. The constant solution, $L'=1/2$ is
physically acceptable and corresponds to the fundamental mode of the string (i.e. to the Theorem \ref{theo1}); for $\alpha>0$
the acceptable solution is $L'_b$, while for $\alpha<0$ the acceptable solution is $L'_d$.

\begin{figure}
\begin{center}
\bigskip\bigskip\bigskip
\includegraphics[width=9cm]{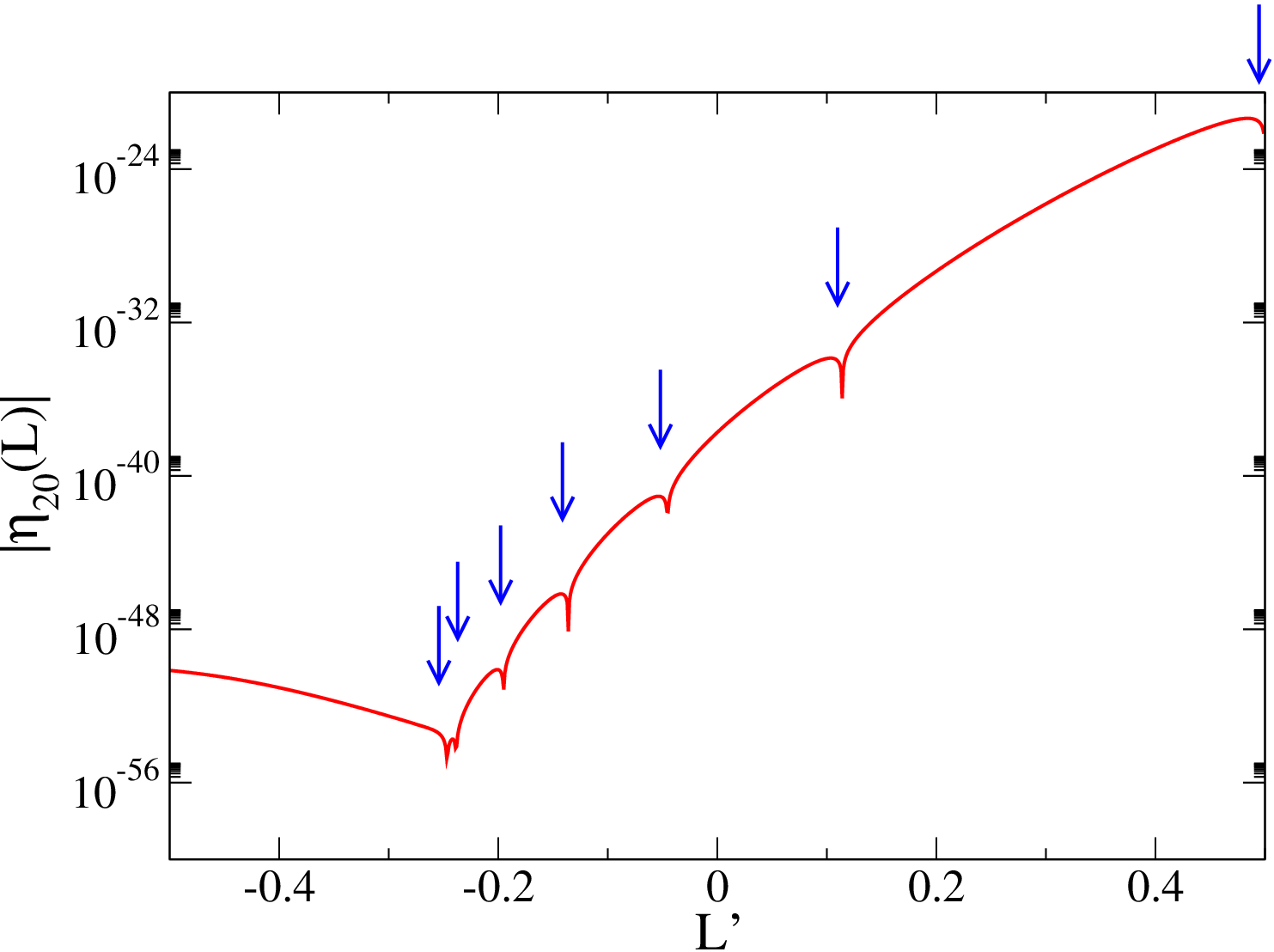}
\caption{$|\eta_{20}(L)|$ as a function of $L'$. The arrows are placed in correspondence of the zeroes of $\eta_{20}(L)$
and locate  approximate solutions of the Helmholtz equation.}
\label{Fig_6}
\end{center}
\end{figure}

Fixing $\alpha=1$ and working through order $20$ we have been able to obtain an explicit expression for $\eta_{20}(x)$: in Fig.~\ref{Fig_6}
where we display $|\eta_{20}(L)|$ as a function of $L'$, we see that there are $7$ zeroes (see the arrows in
the plot) which correspond to an equal number of approximate solutions of eq.~(\ref{helm1d}). Notice that the $\eta_j(x)$ are not
normalized and that we have used a logarithmic scale in the vertical axis (so that the zeroes of $|\eta_n(L)|$ correspond to the 
spikes in the plot).

In Table \ref{table-3} we report the first seven eigenvalues  of the Helmholtz equation (\ref{helm1d}) corresponding to 
a parabolic density with $\alpha=1$ obtained using
Theorem \ref{theo3} with $20$ iterations (third row). The second row reports the values of $L'$ for which $\eta_{20}(L)=0$; the fourth and
fifth rows report the results obtained respectively with collocation with a grid corresponding to $N=2000$ and with the WKB formula. 
Notice that the first $5$ eigenvalues obtained with the Theorem are in almost perfect agreement with the very precise results obtained with 
collocation; the last two eigenvalues on the other hand are rather poor.

At this point we can make few observations:
\begin{itemize}
\item  The number of zeroes of $\eta_{j}(L)$, $j$ being the index of the iterations, grows with $j$: this means
that by iterating a large number of times, one may obtain at once a large number of approximate solutions
of  equation (\ref{helm1d}); these solutions all correspond to the same $\eta_j$ with a different value of $L'$;
\item No orthogonalization is needed, although it could still be used to improve the quality of the results;
\item Using a good ansatz may help reduce the number of iterations:
\item This approach requires that we can perform the integrals in eq.~(\ref{iter3}), 
      which become more and more involved as the iterations are increased;
\end{itemize}

\begin{table}
\begin{center}
\begin{tabular}{|c||c|c|c|c||}
	\hline
$n$	& $L'$  & $E_n$ & $E_n^{CCM}$  &$E_n^{WKB}$ \\  
	\hline
$1$     & $\frac{1}{2}$   & $9.191320572$   &   $9.191320572$  &   $9.869604401$   \\
$2$     & $0.114032016$   & $38.52785042$   &   $38.52785043$  &   $39.47841760$   \\
$3$     & $-0.04546100$   & $87.74309017$   &   $87.74309017$  &   $88.82643961$   \\
$4$     & $-0.135971774$  & $156.75532022$  &   $156.75532022$ &   $157.91367040$  \\
$5$     & $-0.195191018$  & $245.53541494$  &   $245.53541494$ &   $246.74011000$  \\
$6$     & $-0.238496809$  & $354.5795779$   &   $354.0705619$  &   $355.3057584$   \\
$7$     & $-0.246658591$  & $391.6830664$   &   $482.3543658$  &   $483.6106157$   \\
\hline \hline
\end{tabular}
\caption{First seven eigenvalues of the Helmholtz equation (\ref{helm1d}) corresponding to a parabolic density with $\alpha=1$ obtained using
Theorem \ref{theo3} with $20$ iterations (third column). The second column reports the values of $L'$ for which $\eta_{20}(L)=0$; the fourth and
fifth columns report the results obtained respectively with collocation with a grid corresponding to $N=2000$ and with the WKB formula. }
\label{table-3}
\end{center}
\end{table}

\section{Conclusions}
\label{conclusions}

In this paper we have discussed the problem of a string of variable density. 

Our results show that:
\begin{itemize}
\item It is possible to build iteratively precise approximations to the fundamental and excited modes of the string, which
converge to the exact ones as the number of iterations tends to infinity;
\item The results of perturbation theory to second order agree with those of the WKB method for the highly excited modes of the strings,
and imply a functional dependence on the density which is different from the one in two or higher dimensions:  in ref.~\cite{Amore10b}
we have obtained that Weyl's law for a d-dimensional cube of side $2L$ filled with density $\Sigma(x_1,\dots, x_d)$ is
\beq
E_N &\approx& \frac{\pi}{L^2} \frac{(2L)^d \left(\Gamma(d/2+1) \ N \right)^{2/d}}{\int_{\Omega_d} \Sigma(x_1,\dots, x_d)}  \ .
\eeq

\item A collocation method allows to obtain very precise numerical results for large number of modes;
\end{itemize}

\bigskip

The author ackowledges support of Conacyt through the SNI fellowship.


\begin{thebibliography}{}
\bibitem{Krein51}  
M.G. Krein, Determination of the density of a nonhomogeneous symmetric cord by its
frequency spectrum. Doklady Akad. Nauk SSSR (N.S.) 76, 345–348. (1951)
\bibitem{Krein52a} M.G. Krein,  On a generalization of investigations of Stieltjes. Doklady Akad. Nauk SSSR
(N.S.) 87, 881–884. (1952)
\bibitem{Krein52b} 
M.G. Krein, On inverse problems for a nonhomogeneous cord. Doklady Akad. Nauk SSSR
(N.S.) 82, 669–672. (1952b)
\bibitem{Krein53} 
M.G. Krein, On some cases of effective determination of the density of an inhomogeneous
cord from its spectral function. Doklady Akad. Nauk SSSR (N.S.) 93, 617–620. (1953)
\bibitem{Krein54} M.G. Krein, On a method of effective solution of an inverse boundary problem. Doklady
Akad. Nauk SSSR (N.S.) 94, 987–990. (1954)
\bibitem{BSS07} R. Beals, D.H. Sattinger and J. Szmigielski,
The string density problem and the Camassa–Holm equation, Phil. Trans. R. Soc. A {\bf 365}, 2299-2312 (2007)
\bibitem{Rayleigh45} J.W.S. Rayleigh, The Theory of Sound vol 1 (New York: Dover) (1945)
\bibitem{Borg46} G. Borg, Eine Umkehrung der Sturm-Liouvilleschen Eigenwertaufgabe. Acta Math. 78, 1-96. (doi:10.1007/BF02421600) (1946)
\bibitem{Horgan99}  C.O. Horgan and A.M. Chan, 
Vibration of inhomogeneous strings, rods and membranes, Journal of Sound and Vibration {\bf 225}, 503-513 (1999)
\bibitem{Gottlieb02} H.P.W. Gottlieb,  Isospectral strings, Inverse Problems {\bf 18} 971-978 (2002)
\bibitem{Fulcher85} L.P. Fulcher,  Study of the eigenvalues of a nonuniform string, Am.J.Phys.{\bf 53} 730-5 (1985)
\bibitem{BO78}  C.M. Bender and S.A. Orszag, Advanced mathematical methods for scientists and engineers, McGraw-Hill (1978)
\bibitem{Crawford87}  F.S. Crawford,  Eigenvalues of a nonuniform string, Am.J.Phys. {\bf 55}, 168-169 (1987)
\bibitem{CZ00a} C. Castro and E. Zuazua,  High frequency asymptotic analysis of a string with rapidly oscillating density, 
European Journal of Applied Mathematics {\bf 11}, 595-622 (2000)
\bibitem{CZ00b} C. Castro and E. Zuazua, Low Frequency Asymptotic Analysis of a String with Rapidly Oscillating Density,  
SIAM J. Appl. Math. {\bf 60}, 1205-1233 (2000)
\bibitem{RL10} G. Rawitscher and J.Liss, The vibrating inhomogeneous string: a topic for a course in Computational Physics,
arXiv:1006.1913v1 (2010)
\bibitem{Amore10b} P. Amore, Can one hear the density of a drum? Weyl's law for inhomogeneous media, arXiv:0912.1402v1 [math-ph] (2009)
\bibitem{Amore08}  P. Amore, 
Solving the Helmholtz equation for membranes of arbitrary shape: numerical results, 
Journal of Physics {\bf A} 41, 265206 (2008)
\bibitem{Amore10} P. Amore,
Spectroscopy of drums and quantum billiards: perturbative and non perturbative results,
J. Math. Phys. {\bf 51}  (2010)
\bibitem{AA10} C. Alvarado and P. Amore, 
Spectroscopy of annular drums and quantum rings: perturbative and non perturbative results, 
arXiv:1004.2407v1 [quant-ph] (2010)
\bibitem{Amore07} P.Amore et al., 
Variational collocation on finite intervals,
Journal of Physics {\bf A} 40, 13047 (2007)
\bibitem{Amore09}  P. Amore et al.,
Collocation on uniform grids,  Journal of Physics {\bf A} 42, 115302 (2009)











\end{thebibliography}
\end{document}